\documentclass[preprint]{elsarticle}

\makeatletter
\renewcommand\theaffn{\arabic{affn}}
\makeatother

\usepackage{amssymb}

\usepackage[caption=false,font=scriptsize,labelfont=sf,textfont=sf]{subfig}

\usepackage{url}
\usepackage{xurl}
\usepackage{amsmath, amsfonts}
\usepackage{amsthm}
\usepackage{array}
\usepackage{algorithm}
\usepackage{algorithmic}

\usepackage{booktabs}
\usepackage{multirow}
\usepackage{tabularx}
\usepackage{makecell} 

  \graphicspath{{./pdf/}{./jpeg/}{./figures/}}

\usepackage[numbers]{natbib}

\journal{Nature Energy}

\begin{document}

\begin{frontmatter}



  \title{Industrial overcapacity can enable seasonal flexibility in electricity use}


  \author[inst1,inst2,inst3]{Ruike Lyu}
  \author[inst2,inst3]{Anna Li}
  \author[inst4]{Jianxiao Wang}
  \author[inst2]{Hongxi Luo}
  \author[inst1]{Yan Shen}
  \author[inst1,inst7]{Hongye Guo}
  \author[inst5]{Ershun Du}
  \author[inst1,inst6,inst7]{Chongqing Kang}
  \author[inst2,inst3]{Jesse Jenkins}

  \affiliation[inst1]{organization={State Key Laboratory of Power System Operation and Control, Department of Electrical Engineering, Tsinghua University}, city={Beijing}, country={China}}
  \affiliation[inst2]{organization={Andlinger Center for Energy and the Environment, Princeton University}, city={Princeton}, country={USA}}
  \affiliation[inst3]{organization={Department of Mechanical and Aerospace Engineering, Princeton University}, city={Princeton}, country={USA}}
  \affiliation[inst4]{organization={National Engineering Laboratory for Big Data Analysis and Applications, Peking University}, city={Beijing}, country={China}}
  \affiliation[inst5]{organization={Lab of Low Carbon Energy, Tsinghua University}, city={Beijing}, country={China}}
  \affiliation[inst6]{organization={Sichuan Energy Internet Research Institute Tsinghua University}, city={Chengdu}, country={China}}
  \affiliation[inst7]{organization={Corresponding authors}, country={Email: hyguo@tsinghua.edu.cn; cqkang@tsinghua.edu.cn}}

\begin{abstract}
  In many countries, declining demand in energy-intensive industries (EIIs) such as cement, steel, and aluminum is leading to industrial overcapacity. Although industrial overcapacity is traditionally envisioned as problematic and resource-wasteful, it could unlock EIIs' flexibility in electricity use. Here, using China's aluminum smelting industry as a case study, we evaluate the system-level cost-benefit of retaining EII overcapacity for flexible electricity use in decarbonized energy systems. We find that overcapacity can enable aluminum smelters to adopt a seasonal operation paradigm, ceasing production during winter load peaks that are exacerbated by heating electrification and renewable seasonality. This seasonal operation paradigm could reduce the investment and operational costs of China's decarbonized electricity system by 23-32 billion CNY/year (11-15\% of the aluminum smelting industry's product value), sufficient to offset the increased smelter maintenance and product storage costs associated with overcapacity. It may also create labor complementarities between the aluminum and thermal power sectors.
\end{abstract}




\end{frontmatter}



\section*{Introduction}

Industrial restructuring — the transition of economies from heavy industry to higher-added-value sectors~\citep{tan2024toward} — is leading to declining demand for industrial products and overcapacity in energy-intensive industries in many countries. China's steel and cement industries, for instance, have averaged 20-30\% overcapacity in the last decade, resulting in declining profitability and asset utilization~\citep{xie2023impact}. At the same time, the deep decarbonization of these so-called ``hard-to-abate sectors'' requires structural shifts (e.g., increased recycling and electrification-driven technology substitution) that will exacerbate the overcapacity of existing primary (ore-based) production facilities~\citep{nilsson_industrial_2021, meng_technologies_2025, bataille_review_2018, wesseling_transition_2017}. Studies project massive overcapacity for China's aluminum smelters as recycled aluminum increasingly dominates supply (potentially fulfilling 70\% of future demand~\citep{li_when_2020, liu_stock_2013}). Traditionally, overcapacity is viewed as problematic, causing resource waste and economic risks~\citep{yu2020environmental}. Consequently, governments typically strive to eliminate overcapacity through policy measures~\citep{guo2022uncovering}, such as China's Supply-Side Structural Reform (SSSR), which was initiated around 2016 to tackle massive industrial overcapacity.

On the other hand, excess capacity also unlocks operational flexibility for energy-intensive industries, as it frees them from operating at full capacity at all times to meet demand. In decarbonized power systems reliant on variable renewable energy, which are inherently prone to multi-timescale (e.g., both diurnal and seasonal) supply-demand mismatches~\citep{cole_quantifying_2021, denholm_challenges_2021}, the demand-side flexibility of energy-intensive industries may become more valuable for balancing electricity supply and demand \citep{kondziella2016flexibility}. By shifting electricity consumption to periods of high renewable energy availability, energy-intensive industries can reduce both the energy system operational costs and carbon emissions~\citep{golmohamadi_demand-side_2022}. Furthermore, unlike “fast-burst” demand-side resources such as electric vehicles and air conditioning~\citep{van_heerden_demand-side_2025}, industrial facilities with overcapacity and material storage could potentially provide long-duration flexibility by scheduling their production over longer time horizons, functioning as a demand-side alternative to firm power or long-duration energy storage~\citep{sepulveda_role_2018, jenkins_long-duration_2021}. In decarbonized energy systems with increasing variability in electricity prices~\citep{Bistline2021VariabilityID, dardor_modelling_2026}, this flexibility can also lower the energy input costs of energy-intensive industries like aluminum smelting, in which electricity costs exceed 30\% of product costs~\citep{CHALCO2020AnnualReport}.

To quantitatively assess this trade-off, we select China’s aluminum smelting industry as a representative case, because of its strong potential to become a source of flexibility for the power grid. Aluminum smelting is ten times more electricity-intensive per tonne of product than steelmaking~\citep{IAI2023Smelting, Hasanbeigi2019}, and is also likely to have future overcapacity (over 50\% by 2050~\citep{li_when_2020, song_chinas_2023, eheliyagoda2022role, zhang_global_2024}) that allows it to leverage its technical potential for operational flexibility. The aluminum industry is also becoming more integrated with an electricity grid that is increasingly dominated by renewables. Aluminum smelters have historically relied on on-site coal power and were located in northern coal hubs like Shandong~\citep{nakano_decarbonizing_2022}. However, they have recently been relocating to southern provinces like Yunnan, in order to take advantage of low-cost, renewable electricity in these regional grids~\citep{samadi_renewables_2023}. Because of their massive electricity footprint and technical flexibility~\citep{todd2008providing}, these smelters have already been routinely targeted for mandated winter power curtailments during seasonal hydropower droughts~\citep{alcircle_yunnan_cuts_2024}. Given these recent trends, aluminum smelting is an ideal archetype for examining the system-level benefits of overcapacity-enabled industrial flexibility.

Traditionally, demand-side flexibility has been viewed primarily as the ability to curtail or shift load for a few hours during extreme price spikes. However, deeply-decarbonized energy systems will also require long-term flexibility due to seasonal mismatch between electricity supply and demand. While extensive research has focused on supply-side seasonal balancing (e.g., long-duration thermal storage and hydrogen), long-duration flexibility on the demand side has received little attention. In the case of aluminum smelting, for instance, existing studies have primarily focused on validating the technical feasibility and potline-level thermal constraints of hourly-level flexible operation for short-term demand response~\citep{todd2008providing, wong_studies_2023, tabereaux_loss_2016, driscoll_economics_2016}. To the best of our knowledge, the feasibility and system-level value of long-term flexibility in smelting electricity use remain largely unexplored in current literature.

\begin{figure}[!t]
  \centering
  \includegraphics[width=0.75\textwidth]{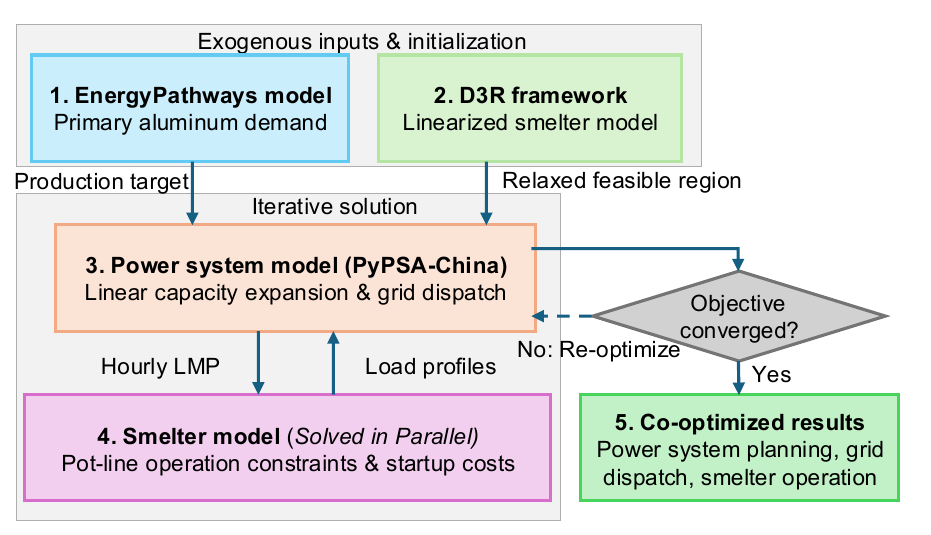}
  \caption{\textbf{Schematic diagram of the integrated power-aluminum co-optimization framework.} To overcome the computational intractability of an 8760-hour horizon, the PyPSA-based grid model and detailed smelter operations are decoupled and solved iteratively through the exchange of local marginal prices (LMPs) and industrial load profiles, aided by the Data-Driven Dimension Reduction (D3R) framework.}
  \label{fig_diagram}
\end{figure}

In this paper, we aim to understand the value of aluminum smelting overcapacity for providing demand flexibility in China's decarbonized energy system. We examine these effects across both physical infrastructure and socio-economic dimensions (e.g., labor deployment), as both are essential for charting a sustainable path for the aluminum industry. However, the computational complexity of modeling and optimizing the intricate coupling between two already large and complex systems—the electricity system and aluminum smelters— is a significant barrier to such analysis and has not been done before.
We developed an integrated provincial-level power-smelting co-optimization model (Figure \ref{fig_diagram}). This approach couples the PyPSA power system framework~\citep{PyPSA} with a detailed aluminum smelter operation model~\citep{shen_improved_2025} for hourly co-optimization. To solve this computationally heavy 8760-hour simulation, we adopted a Data-Driven Dimension Reduction (D3R) framework~\citep{lyu_data-driven_2025} paired with an iterative decomposition algorithm. First, D3R linearizes complex smelter constraints and generates an initial co-optimized solution to make the power system model easier to solve. The power system model then passes hourly locational marginal prices (LMPs) to the smelters. Using these price signals, each smelter independently optimizes its discrete production decisions (e.g., shutdown and restart costs) in parallel. This price-load feedback loop iterates until convergence, bridging modeling high-fidelity smelting physics with large-scale grid planning (see Methods for details).

We find that in decarbonized electricity systems, overcapacity enables aluminum smelters to shift operations away from periods of winter electricity supply scarcity, which are driven by low renewable availability and heating electrification. This flexibility lowers overall electricity system costs by reducing peak loads, which in turn decreases the generation capacity required to meet those peaks, as well as lowering utilization of coal, natural gas, and nuclear generation. In the Decarbonized Power 2050 (DP-2050) scenario, retaining 30\% aluminum-smelting overcapacity reduces electricity system costs by 23-32 billion CNY/year (or 11-15\% of the aluminum smelting industry's product value), as compared to a no-overcapacity case. We find that, in such a scenario, aluminum production costs are lower by over 1,500 CNY/tonne (or 9\% of the total production cost), due to aluminum smelters consuming cheaper electricity.
We also explore the potential socio-economic implications of this transition. Because thermal power plants may also operate seasonally in decarbonized systems, complementary seasonal operation in the aluminum smelting industry reveals a potential for inter-industry labor deployment. Our analysis indicates that if workers could switch between industries on a seasonal basis, aggregated workforce volatility across the aluminum smelting and power generation sectors could theoretically be reduced by up to 25\%. This highlights a possible co-benefit of seasonal aluminum smelter operation for bolstering social stability in deeply-decarbonized energy systems.

Although we are using aluminum smelting as a case, our findings can be generalized to other energy-intensive industries, such as steel and cement industries, where seasonal operation could potentially bring significant energy input cost savings to offset the costs of maintaining overcapacity and product storage. More broadly, our work suggests that industrial overcapacity could be reframed from a problem to an opportunity to reduce energy system costs, lower industrial production costs, foster workforce stability, and thereby mitigate socio-economic disruptions from energy decarbonization.

\subsection*{Overcapacity in aluminum smelting in the recycle era}

\begin{table}[htbp]
  \centering
  \caption{Outline of Model Scenario Settings.}
  \label{tab:scenario_settings}
  \footnotesize 
  \setlength{\tabcolsep}{3pt} 
  \begin{tabularx}{\linewidth}{>{\raggedright\arraybackslash}X >{\raggedright\arraybackslash}X >{\raggedright\arraybackslash}X >{\raggedright\arraybackslash}X >{\raggedright\arraybackslash}X}
  \hline
  \multicolumn{5}{c}{Primary aluminum demand} \\ \hline
  \multirow{2}{*}{Case} & \multirow{2}{*}{\makecell{Stock \\saturation  level\\(kg/capita)}} & \multirow{2}{*}{\makecell{Scrap import \\ rate}} & \multirow{2}{*}{\makecell{Scrap \\recycle rate}} & \multirow{2}{*}{\makecell{Product mean \\ lifetime \\(year)}} \\
  & & & & \\ \\ \hline
  \multirow{2}{*}{Low} & \multirow{2}{*}{\makecell{500}} & \multirow{2}{*}{10\%} & \multirow{2}{*}{75\%} & \multirow{2}{*}{18} \\
  & & & & \\
  \multirow{2}{*}{Mid} & \multirow{2}{*}{\makecell{600}} & \multirow{2}{*}{5\%} & \multirow{2}{*}{70\%} & \multirow{2}{*}{16} \\
  & & & & \\
  \multirow{2}{*}{High} & \multirow{2}{*}{\makecell{700}} & \multirow{2}{*}{0\%} & \multirow{2}{*}{65\%} & \multirow{2}{*}{14} \\
  & & & & \\ \hline
  \multicolumn{5}{c}{Smelter operational flexibility} \\ \hline
  \multirow{2}{*}{Case} & \multirow{2}{*}{\makecell[l]{Min. power consumption\\(per unit of nameplate value)}} & \multirow{2}{*}{\makecell[l]{}} & \multirow{2}{*}{\makecell[l]{Restart costs\\(EUR/MW)}} & \multirow{2}{*}{\makecell[l]{}}  \\
  & & & & \\ \hline
  \multirow{2}{*}{Low} & \multirow{2}{*}{$0.99$} & \multirow{2}{*}{}& \multirow{2}{*}{96,594}  & \multirow{2}{*}{} \\
  & & & & \\
  \multirow{2}{*}{Mid} & \multirow{2}{*}{$0.9$} & \multirow{2}{*}{} & \multirow{2}{*}{13,981}  & \multirow{2}{*}{} \\
  & & & & \\
  \multirow{2}{*}{High} & \multirow{2}{*}{$0.7$} & \multirow{2}{*}{} & \multirow{2}{*}{2,796}  & \multirow{2}{*}{} \\
  & & & & \\
  \multirow{2}{*}{Unconstrained} & \multirow{2}{*}{$0.0$} & \multirow{2}{*}{} & \multirow{2}{*}{0}  & \multirow{2}{*}{} \\
  & & & & \\ \hline
  \multicolumn{5}{c}{\makecell{Technology CAPEX}} \\ \hline
  \multirow{2}{*}{Case} & \multirow{2}{*}{\makecell{Wind/solar\\(EUR/kW$_{el}$)}} & \multirow{2}{*}{\makecell{Battery\\(EUR/kWh)}} & \multirow{2}{*}{\scriptsize \makecell{Long-duration\\storage\\(EUR/kW$_{el}$)}} & \multirow{2}{*}{\scriptsize \makecell{Methanation/\\DAC\\(EUR/kW$_{CH_4}$)}} \\ \\
  & & & & \\ \hline
  \multirow{2}{*}{Low} & \multirow{2}{*}{624/346} & \multirow{2}{*}{186} & \multirow{2}{*}{520} & \multirow{2}{*}{575} \\
  & & & & \\
  \multirow{2}{*}{Mid} & \multirow{2}{*}{780/432} & \multirow{2}{*}{232} & \multirow{2}{*}{650} & \multirow{2}{*}{719} \\
  & & & & \\
  \multirow{2}{*}{High} & \multirow{2}{*}{780/432} & \multirow{2}{*}{348} & \multirow{2}{*}{975} & \multirow{2}{*}{1079} \\
  & & & & \\ \hline
  \multicolumn{5}{c}{Labor and employment flexibility} \\ \hline
  Case & \multicolumn{2}{l}{Labor and standby costs} & \multicolumn{2}{l}{Workforce mobility} \\ \hline
  \multirow{2}{*}{Inflexible} & \multicolumn{2}{l}{Constant regardless of output} & \multicolumn{2}{l}{None (standby in smelters)} \\
  & \multicolumn{2}{l}{} & \multicolumn{2}{l}{} \\
  \multirow{2}{*}{Flexible} & \multicolumn{2}{l}{Reduced during shutdown} & \multicolumn{2}{l}{Seasonal transfer to other sectors} \\
  & \multicolumn{2}{l}{} & \multicolumn{2}{l}{} \\ \hline
  \end{tabularx}
  
  \raggedright
  \footnotesize
  The restart-up costs are measured from the electricity input side. Exchange rates throughout the study (annual averages 2020): 1\,USD = 6.900\,CNY; 1\,EUR = 7.868\,CNY; 1\,USD = 0.877\,EUR. The electricity system cost data is primarily from the PyPSA-China model~\citep{zhou_multienergy_2024}, with the specific Chinese cost assumptions and their original sources documented in Supplementary Table 7. Long-duration storage includes H2 and pumped hydro storage. Our core scenario is defined by the combination of Mid primary aluminum demand, Mid smelter operational flexibility, Mid technology costs, and Inflexible labor. 
  \normalsize
\end{table}

This study employs a scenario analysis approach to explore the impact of key uncertainties on the system-level value of aluminum smelting overcapacity. As detailed in Table \ref{tab:scenario_settings}, we evaluate four primary dimensions: (i) primary aluminum demand, which determines the industry's aggregate flexibility potential by defining the size of the sector; (ii) smelter operational flexibility, representing the physical minimum load constraints and the economic penalty (e.g., restart costs) of non-baseload operation; (iii) technology costs, reflecting the relative competitiveness of demand-side flexibility against supply-side flexibility alternatives (e.g., batteries and hydrogen); and (iv) labor flexibility, which dictates whether workforce costs can be mitigated during seasonal shutdowns. (see Methods for more details)
To ensure the robustness and traceability of these scenario settings, the bounds for each dimension are systematically derived from empirical literature and industry standards. For technology costs, the Mid scenario adopts peer-reviewed trajectories from the PyPSA-China\citep{zhou_multienergy_2024} dataset, while the Low and High bounds for the costs are scaled by $-20$\% and $+50$\%, respectively, adhering to the AACE International Class 4 estimate guidelines \citep{bates2005cost}.

For each combination of the four scenario dimensions in Table~\ref{tab:scenario_settings} (primary aluminum demand, smelter operational flexibility, technology costs, and labor flexibility)—yielding 72 scenario combinations (3$\times$4$\times$3$\times$2)—we vary the level of retained overcapacity in discrete steps (from 0\% to 100\% of the initial excess capacity) to identify the overcapacity level that maximizes net system benefit. The 0\% retention level serves as our no-overcapacity baseline case, for which excess capacity is assumed to be fully decommissioned, leaving the sector with exactly enough capacity to meet the projected annual demand under continuous, inflexible baseload operation. Parameterizing the overcapacity level over these 72 scenarios leads to over 1000 optimization results, the distribution of which is reported in subsequent figures. Our core scenario throughout this analysis corresponds to the Mid--Mid--Mid--Inflexible combination in Table~\ref{tab:scenario_settings}, and is used to show results in Figures 4a, 5, and 6. Detailed parameter values, data sources, and the comprehensive rationales for the Low, Mid, and High cases (and both Inflexible and Flexible labor cases) across all dimensions are documented in the Methods and Supplementary Notes.

\begin{figure}[!t]
\centering
\includegraphics[width=0.98\textwidth]{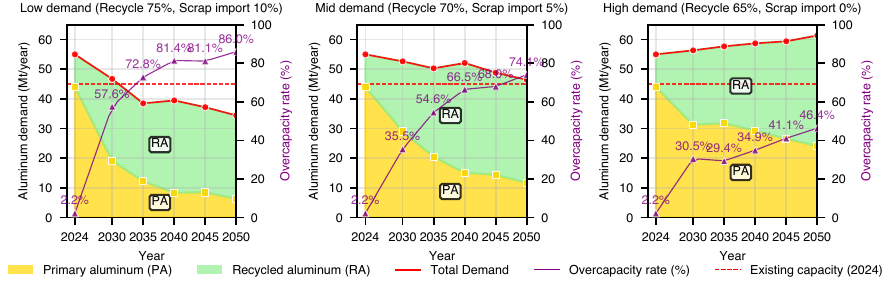}
\caption{\textbf{China's projected aluminum demand and smelting overcapacity under different scenarios.} The top solid line represents total aluminum demand, the green area shows recycled aluminum production (RA), and the yellow area indicates primary aluminum production (PA) up to 2050. China's aluminum smelting capacity at the end of 2024 (45 million tonnes (Mt) per year) is represented by the red dashed line. The area between the red dashed line (smelter capacity at 2024 level) and the yellow area represents the excess capacity of aluminum smelters unless prematurely decommissioned. Because smelters typically operate for over 50 years, the vast majority of China's existing capacity will not reach its natural retirement age before 2050 (see Supplementary Note 3.3), effectively locking in this structural baseline. The overcapacity rate is represented by the purple line, and is given by one minus the utilization rate (1 - demand/capacity). Different scenarios are constructed based on various assumptions regarding economic development and aluminum product import/export levels, with the variable settings outlined in Table \ref{tab:scenario_settings}. Even under the high-demand scenario, China's aluminum industry would retain 46\% excess capacity — and as much as 86\% under the low-demand scenario — posing a significant challenge to the sector.}
\label{fig_demand}
\end{figure}

Our analysis shows that over the next one to two decades, China's primary aluminum demand will rapidly decline and remain low as more aluminum is produced from scrap, leading to unprecedented overcapacity in the aluminum smelting industry. As shown in Figure \ref{fig_demand}, the proportion of recycled (secondary) aluminum production in total aluminum demand will increase from 25\% in 2024~\citep{ChinaNonferrous2025} to 60\% by 2050. The increased recycled aluminum production stems from a significant increase in available scrap aluminum, which is a delayed effect of China's climbing aluminum production over the past two decades, as aluminum products typically have a lifespan of 10-20 years~\citep{hatayama2012evolution}. While this transition entails substantial reductions in the aluminum industry's energy consumption and carbon emissions, due to the lower energy intensity of producing aluminum from scrap~\citep{li_analysis_2021}, it would result in over 60\% excess capacity for existing aluminum smelters. Furthermore, because aluminum smelters typically operate for over 50 years, the vast majority of China's existing smelting capacity will be far from the retirement age, locking in this severe overcapacity (see Supplementary Note 3.3 for the detailed capacity age distribution).

\subsection*{Electricity system savings through flexible smelting}

\begin{figure}[!t]
\centering
\includegraphics[width=1.0\textwidth]{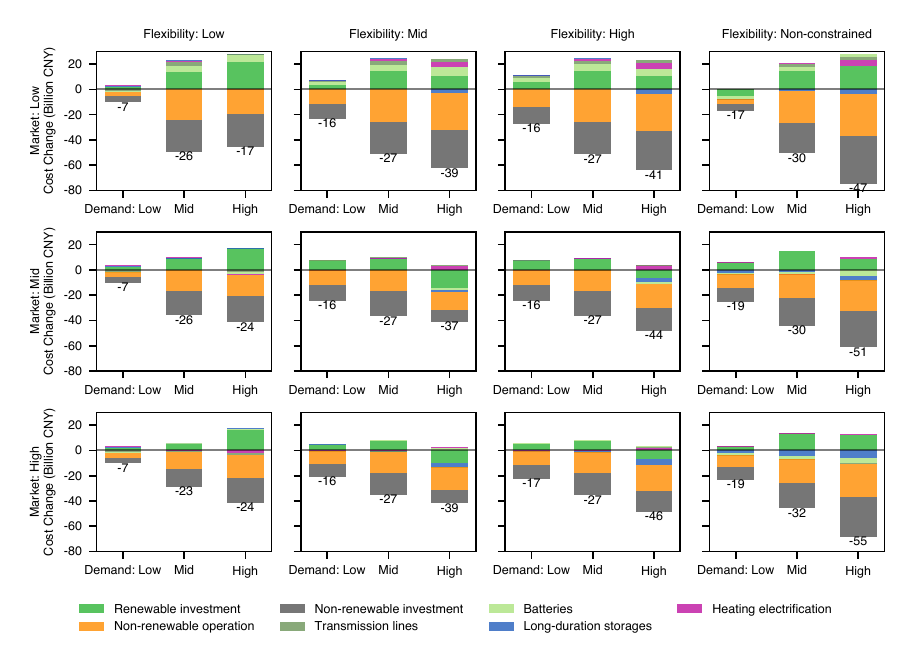}
\caption{\textbf{Reduction in electricity system costs due to overcapacity-enabled flexibility in aluminum smelting, compared to the no-overcapacity case.} Colors represent different sources of changes in system cost: non-renewable investment costs (gray) and non-renewable operational costs (orange) account for the largest system cost reductions, reflecting a decreased reliance on firm generation resources (coal, gas, biomass, and nuclear) to ensure power supply reliability. In Mid-demand scenarios, cost reductions range from 23 to 32 billion CNY (11-15\% of the aluminum smelting industry's product value in 2050), across different smelter flexibility and technology cost settings. Conversely, the observed increases in renewable energy and heating electrification investments indicate that seasonal load-shifting makes these clean technologies more economically efficient. By concentrating electricity consumption during summer periods of abundant renewable generation and avoiding winter peak loads, smelter flexibility eliminates the need for expensive firm generation and massive storage. This allows the system to cost-effectively expand renewable capacity and clean heating without facing severe integration barriers.}
\label{fig_value}
\end{figure}

Figure \ref{fig_value} shows the reduction in electricity system costs in the DP-2050 scenario when the operation of aluminum smelters with overcapacity is co-optimized with the electric power system, as compared to the no-overcapacity case. This represents the value of flexibility in smelting electricity use. Here, the power system cost includes operational costs and the annualized capital investment for new generation, storage, and transmission capacity. The costs of the aluminum smelters are not included in the electricity system costs, but are discussed later in the paper. The results show that in the Mid-demand scenario, the overcapacity of aluminum smelters provides a flexibility value of 23 to 32 billion CNY to the power system. To put this number into perspective, this is equivalent to 11-15\% of the aluminum smelting industry's product value in 2050.

Across all scenarios, the cost reductions are driven by substantial decreases in both the investment and operational costs (mainly fuel costs) of non-renewable energy sources (coal, natural gas, and nuclear). This is accompanied by increased investments in renewable energy and heating electrification in order to reach decarbonization constraints. Two interconnected mechanisms explain this shift toward a more renewable-dominated portfolio. First, without adequate long-duration demand-side flexibility, the expansion of variable wind and solar power faces diminishing marginal benefits, as their high variability and mismatch with electricity consumption lead to severe curtailment or the need for massive storage. By concentrating electricity consumption during periods of high renewable generation in the summer, flexible smelter operation enhances the utilization efficiency of renewables, making it economically beneficial to deploy greater renewable capacity. Second, shifting smelting load to the summer mitigates winter peak loads, significantly reducing the need for costly clean-firm energy (dispatchable low-carbon resources such as coal or natural gas with CCS, biomass with CCS, or fossil power offset by direct air capture~\citep{sepulveda_role_2018}) during periods of renewable scarcity.

Comparing the different scenarios, we find that the flexibility value of aluminum smelting is higher in scenarios with high aluminum demand, because the aluminum smelting sector is bigger and the impact of load flexibility is therefore larger. The overall electricity system cost savings are similar between the Non-constrained-flexibility scenario and the Mid-flexibility scenario, suggesting diminishing returns in increasing smelter operational flexibility.

\subsection*{Cost-benefit analysis of retaining smelting overcapacity}

\begin{figure}[!t]
  \centering
  \subfloat[\scriptsize Annual net benefit of different aluminum smelting overcapacities in 2050\label{fig_tradeoff}]{
    \includegraphics[width=0.47\textwidth]{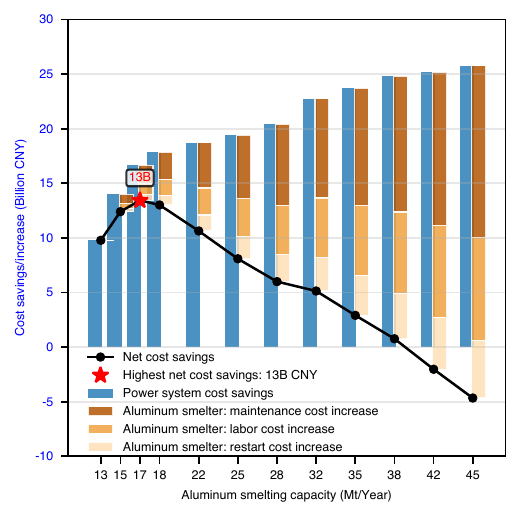}
  }
  \hfill
  \subfloat[\scriptsize Distribution of optimal retained capacity and net benefit across the scenarios\label{fig_optimal}]{
    \includegraphics[width=0.47\textwidth]{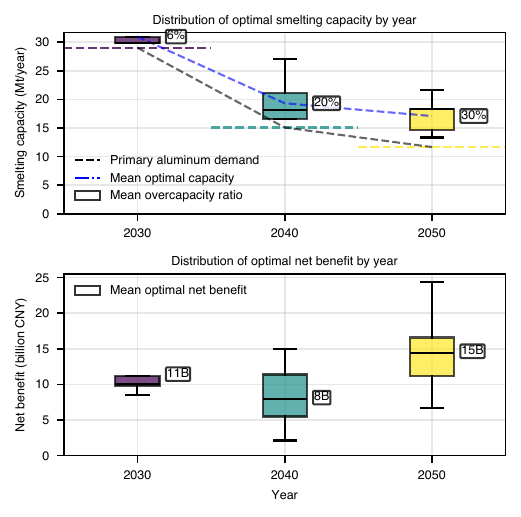}
  }
  \caption{\textbf{Trade-off analysis and cost-effective strategy for retaining overcapacity.} In (a), electricity system cost savings, smelter cost increase, and net system benefit with different retained capacities in 2050 in the core scenario are shown. The maximal net benefit (electricity system cost reduction minus smelter increased costs) is achieved with 30\% overcapacity in 2050. In (b), the mean optimal overcapacity rate increases from 6\% in 2030 to 30\% in 2050, due to higher flexibility requirements as the electricity system approaches net-zero emissions. 
  Notably, while the optimal rate of overcapacity rises, the absolute cost-effective smelting capacity decreases over time due to declining primary aluminum demand. The net benefit across different scenarios first decreases as primary aluminum demand declines to 2040 levels, then increases in the net-zero emissions scenario in 2050, where load flexibility is more valuable.
  All statistics are derived from $n = 36$ independent scenarios per planning year. For each box plot, the center horizontal line represents the median, while the top and bottom box limits indicate the third and first quartiles, respectively.}
  \label{fig_tradeoff_optimal}
\end{figure}

While flexible aluminum smelting with overcapacity can reduce power system costs, it also entails higher maintenance costs, labor cost, restart cost, and extra expenditure, such as product storage. As shown in Figure~\ref{fig_tradeoff_optimal}(a), the reduced power system costs increase with overcapacity but plateau after smelter capacity exceeds 35 Mt/year (60\% overcapacity), while operational costs increase approximately proportionally with excess capacity. A smelting capacity of 17 Mt/year (30\% overcapacity) is found to maximize the net system cost reduction at 13 billion CNY, where the marginal decrease in power system cost equals the marginal increase in smelter operational cost. Retaining more excess capacity could be more attractive if the cost of retiring capacity is taken into account (dismantling retired electrolytic cells, compensating affected smelters, and relocating displaced industrial workers, etc.). These additional costs could further justify maintaining higher levels of excess capacity beyond what our results suggest.

Figure \ref{fig_tradeoff_optimal}(b) shows the optimal smelter capacity and corresponding net system cost reduction for different years and scenarios. The average optimal overcapacity rates across all scenarios are 6\%, 20\%, and 30\% for 2030, 2040, and 2050, respectively. The corresponding net value first decreases from 2030 to 2040, as the declined primary aluminum demand leads to a shrinkage of the aluminum smelting industry, then increases by 2050, approaching net-zero emission targets.

\subsection*{Seasonal operation paradigm enabled by overcapacity}

\begin{figure}[!t]
\centering
\subfloat[\footnotesize Normalized hourly aluminum smelting power consumption and stored aluminum with 30\% overcapacity in 2050 core scenario.\label{fig_operation}]{
  \includegraphics[width=0.8\textwidth]{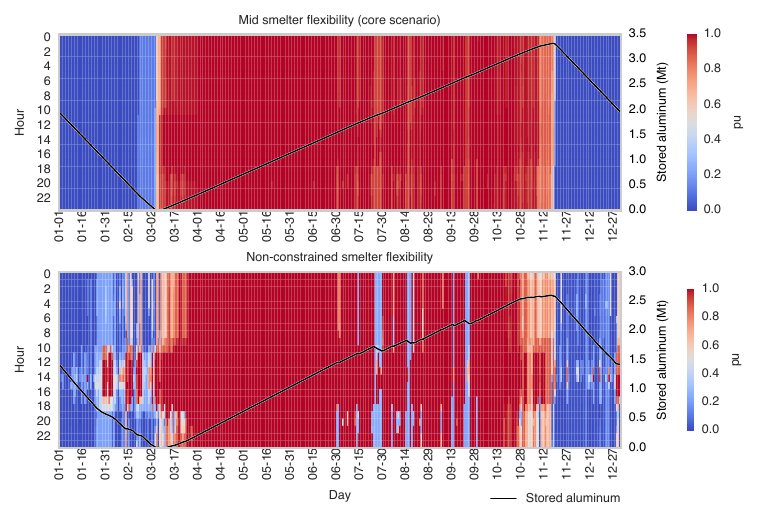}
}
\vspace{0.5cm}
\subfloat[\footnotesize Monthly electricity demand and generation in 2050. (Residual demand = electricity demand $-$ electricity generation)\label{fig_monthly_demand}]{
  \includegraphics[width=0.75\textwidth]{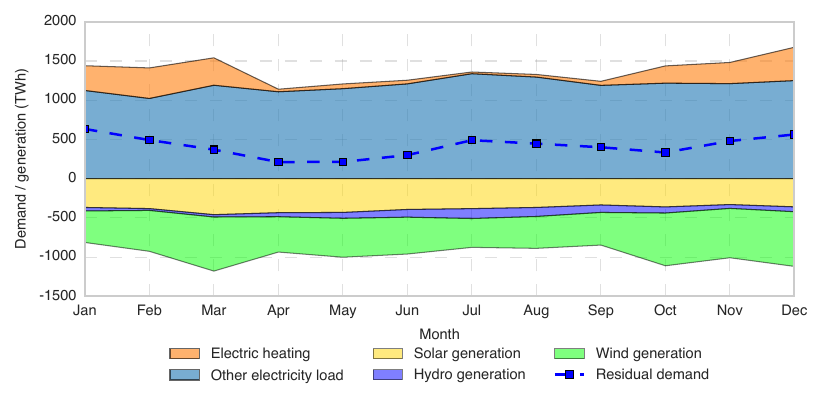}
}
\caption{\textbf{Seasonal operation of aluminum smelters is complementary to energy system seasonality.} In (a), smelter production largely ceases from mid-November to mid-March with demand met by stored inventory, but operates at full capacity from late March to early November. Comparing (a) with (b), the electricity consumption of aluminum smelting is largely complementary to China's monthly residual load, acting as a crucial seasonal flexibility resource to mitigate the winter electricity demand peaks driven by heating electrification.}
\label{fig_operation_demand}
\end{figure}

Figure \ref{fig_operation_demand} illustrates how the seasonal nature of renewable energy generation and heating electrification significantly alters the operational dynamics of aluminum smelters, which adjust their production schedules to align with the seasonality of the energy system. Panel (a) reveals clear seasonal operational patterns: production largely ceases from mid-November to mid-March (the winter heating season), with demand met by inventory, while operating at or near rated capacity from late March to early November. The lower panel shows that even with unconstrained aluminum smelter flexibility, seasonal operational patterns remain dominant in aluminum smelting. Although China faces significant cooling demand in July and August, smelters do not operate flexibly to balance that demand because air conditioning load appears as short-duration peaks that can more cost-effectively be balanced by batteries or vehicle-to-grid (V2G) technologies.

Panel (b) reveals the underlying system driver for this behavior: the emergence of a ``seasonal duck curve''. Driven by high renewable penetration and heating electrification, China's residual demand (electricity demand minus electricity generation) exhibits a distinct and prolonged winter peak. During these winter periods, renewable energy output, especially from hydropower, is at its minimum while heating demand peaks. Although wind power also increases in the winter, it is outpaced by the surge in heating demand, forcing the power system to rely more heavily on gas and coal generation to ensure power supply reliability. Here, the national aggregate solar output obscures regional differences (e.g., reduced summer solar generation in Yunnan due to the monsoon season). For provincial results, see Supplementary Figure 8.

\begin{figure}[!t]
\centering
\includegraphics{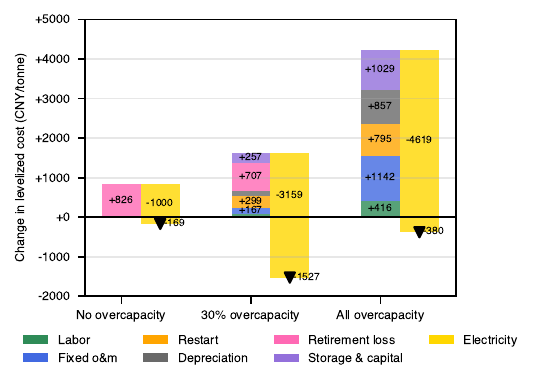}
\caption{\textbf{Comparison of the change in levelized cost per tonne of aluminum between the 2020 baseline and various 2050 overcapacity levels.} In the core scenario, maintaining 30\% overcapacity is beneficial for the smelters, reducing aluminum production costs by over 1,500 CNY/tonne (approximately 9\% of current production costs), with the drop in electricity bills offsetting the maintenance costs of overcapacity and increased restart, storage, and capital costs.}
\label{fig_cost_reduction}
\end{figure}

This results in significantly higher electricity costs in winter, because the system must deploy more direct air capture and/or bioenergy with CCS to offset the carbon emissions. To balance this seasonal supply-demand mismatch and flatten the prolonged winter peak, the system requires dedicated seasonal flexibility resources. Aluminum smelting with overcapacity perfectly fills this role, creating strong economic incentives for smelters to avoid production during these high-cost periods and instead rely on inventory to meet demand. Interestingly, the averaged optimal overcapacity rate of 30\% roughly corresponds to the proportion of winter time in a year, further confirming that this seasonal duck curve, rather than daily fluctuations, dictates the investment logic of overcapacity. 

This seasonal operational paradigm could decrease the cost of producing aluminum (Figure \ref{fig_cost_reduction}). {Our analysis indicates that maintaining a moderate level of overcapacity (e.g., 30\%, middle bar of Figure \ref{fig_cost_reduction}) minimizes overall production costs, outperforming both the complete decommissioning (leftmost bar of Figure \ref{fig_cost_reduction}) and full retention (rightmost bar of Figure \ref{fig_cost_reduction}) scenarios. By retaining 30\% overcapacity, smelters can effectively utilize zero-marginal-cost renewable energy during high-generation periods. The resulting electricity cost savings are sufficient to offset the added expenses of overcapacity maintenance, incremental capital (see Supplementary Note 3.6), and product storage, ultimately reducing total costs by 9\% (over 1,500 CNY/tonne of aluminum). In contrast, if no overcapacity is retained, the savings from general renewable penetration are largely negated by equipment depreciation and early decommissioning costs. Conversely, retaining 100\% of the initial overcapacity introduces fixed maintenance and capital burdens that outweigh the benefits of seasonal flexibility in electricity use.}

The reduced aluminum production costs, along with the electricity system cost savings shown previously, demonstrate that the seasonal flexibility strategy creates a win-win scenario for both the grid and aluminum producers, altering the economic viability of maintaining excess capacity. It is worth noting that our current model only captures savings from electricity procurement. As electricity market designs evolve to value system adequacy, industrial facilities operating seasonally will likely capture additional value through capacity markets. Consequently, the actual financial returns for aluminum smelters are expected to be even more optimistic than our current projections. The detailed cost composition underlying these figures is provided in Supplementary Note 3.4.

\begin{figure}[!t]
  \centering
  \includegraphics[width=1.0\textwidth]{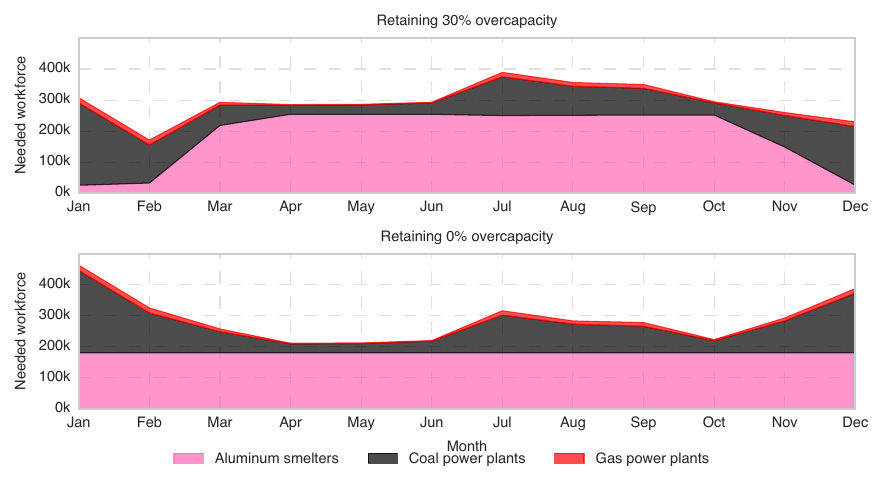}
  \caption{\textbf{Estimated monthly workforce requirements for aluminum smelters and coal/gas power plants in 2050 with and without aluminum smelting overcapacity.} The results are from the flexible labor scenario with Mid-primary aluminum demand, Mid-smelter flexibility, and Mid-power system technology cost. With 30\% smelter overcapacity, there is a potential temporal complementarity of labor demand between smelters and thermal generators. If such potential is realized, complementary employment patterns between the two sectors could reduce workforce variability, further unleashing the social benefits of seasonal flexibility.}
  \label{fig_operation_employment}
\end{figure}

As Figure \ref{fig_operation_employment} illustrates, the seasonally-operating aluminum smelting industry and the power sector have complementary workforce demand patterns, with coal and gas power plants needing more workers in the winter, and the aluminum smelting industry needing more workers in spring, summer, and fall. Because China's largest aluminum companies have historically operated their own captive power plants (see Supplementary Note 3.8), there exists a possible scenario in which a subset of China's aluminum companies move workers between aluminum smelters and power plants from season to season~\citep{tan_different_2025}. This workforce management paradigm could reduce total seasonal employment fluctuation (calculated as the standard deviation of monthly employment throughout the year) across the two sectors by 25\%. Maintaining a larger capacity also increases overall employment compared to the no-overcapacity scenario. While implementing inter-industry labor mobility depends on factors such as worker training and flexible employment relations, this coupling could potentially mitigate social disruption from job losses that may result from both energy system transitions and industrial restructuring, highlighting the potential societal value of overcapacity-enabled seasonal flexibility.

\section*{Discussion}\label{sec_discussion}

In power systems with increasing penetration of renewable energy, the volatility of electricity supply is often compounded by load fluctuations from heating electrification, creating significant seasonal variations in electricity supply and demand. Concurrently, many traditional energy-intensive industries face persistent overcapacity, which reduces capacity utilization and creates uncertainty on their profitability. This convergence of challenges makes it increasingly urgent to understand how energy-intensive industries can adapt and thrive in decarbonized energy systems.

In this work, we use China's aluminum smelting industry—a highly energy-intensive sector facing unprecedented overcapacity in the coming recycling era—as a representative case to explore the flexibility value of overcapacity in decarbonized energy systems. We developed a comprehensive framework that co-optimizes power system capacity expansion planning with detailed smelter operations, capturing key technical constraints such as temperature limits, restart costs, and operational limitations. This co-optimization was evaluated across more than 1000 scenarios, varying flexibility settings and competing technology costs, to inform cost-effective strategies for retaining overcapacity and leveraging it for seasonal flexibility.

Our findings reveal that overcapacity unlocks the flexibility of aluminum smelting operation, allowing smelters to curtail production during the winter when peak demand is exacerbated by heating electrification and reduced renewable output. We identify an optimal overcapacity rate of approximately 30\% for the aluminum smelting industry in China's decarbonized electricity power system, which closely matches the fraction of the year that is the winter heating season, indicating that the seasonality of the energy system translates into seasonal electricity price signals, which ultimately drive the smelters' seasonal operation. This seasonal operation transforms a flat industrial power demand into a seasonal load profile that reduces mismatch between electricity supply and demand. This avoids the need for massive investment and operational costs for expensive clean-firm power technologies, while simultaneously improving renewable energy integration. These reduced investment and operational costs translate into cheaper electricity costs, sufficient to offset the maintenance and operational costs of excess capacity and product storage, thereby cutting aluminum production costs by approximately 9\% of the current level.

Crucially, our findings suggest a need to re-evaluate the traditional paradigm of demand response, which in existing literature predominantly targets short-term (hourly to daily) demand-side flexibility. We demonstrate that energy-intensive industrial sectors have strong potential to be advantageous for long-term, seasonal demand flexibility. Industrial processes like primary aluminum smelting are fundamentally designed for steady-state, continuous operation. Forcing these systems to provide high-frequency, short-term flexibility often incurs steep efficiency penalties, thermal fatigue of equipment, risks of cell shutdown, and high control costs. In contrast, leveraging overcapacity allows these facilities to operate continuously at optimal efficiency for most of the year, undergoing only planned, long-duration curtailments during extreme seasonal grid peaks. Therefore, when accounting for the strict requirements of stable industrial operations, this seasonal paradigm emerges as not only economically superior but also technically far more practical than conventional short-term demand response.

More broadly, our work suggests that industrial overcapacity should be reframed from a problem to an opportunity in a world where low-cost but variable renewable energy is becoming one of the cheapest and most abundant forms of primary energy available for industrial activity. Just as the global economy has reconfigured multiple times around the availability and characteristics of new low-cost primary energy sources, the ongoing shift to renewable electricity may entail new paradigms for industrial operations, plant design, and the geography of primary industries. While our work focuses on aluminum smelting as a case study, these findings are potentially generalizable to other electricity-intensive industries facing overcapacity, such as steel and cement. By embracing this new seasonally-operating paradigm, these sectors may also be able to leverage overcapacity and product storage to adapt to volatile seasonal electricity prices, reducing both their own operational expenses and overall energy system costs, while enabling more cost-effective mitigation of greenhouse gas emissions.

Realizing this seasonal paradigm also requires addressing its socio-economic implications, particularly concerning labor. However, because the labor intensity of modern primary aluminum production is exceptionally low (accounting for only 2\%--3\% of total costs~\citep{smm_costs_2024}), the immense energy cost savings achieved during high-generation periods are more than sufficient to cover standby wages or training stipends during winter lulls. As financial incentives make the transition to seasonal operations economically compelling, exploring potential employment complementarity under this paradigm represents a promising avenue for future research and policy design. For instance, the winter production lull in aluminum smelting naturally coincides with the operational peaks of thermal power plants in a decarbonized grid. While facilitating such inter-industry job mobility requires specific spatial planning (e.g., large integrated industrial parks) and targeted institutional support, it is well-grounded in China's empirical reality. Formalized national policies such as the ``shared employee'' framework, alongside the internal labor markets of large state-owned conglomerates, already facilitate similar seasonal workforce rotations across manufacturing and heavy industries (see Supplementary Notes 6.1). Ultimately, supported by carefully designed market and policy mechanisms, the seasonal operation of energy-intensive industries holds the potential to unlock broader societal benefits, transforming structural overcapacity into a strategic asset for employment resilience and social stability.

\section*{Methods}\label{sec_methods}

\subsection*{Data Source and Scenario Settings}\label{sec:data_source_and_scenario_settings}

The factors that have significant and uncertain impacts on the value of overcapacity in the aluminum industry include: the demand for primary aluminum (from smelting), the operational flexibility of the smelters, the costs of other low-carbon technologies for load flexibility (batteries, long-duration storage, etc.), and the flexibility of labor and employment. This study employs a scenario analysis approach, grouping these factors into four-dimensional parameter sets (Table \ref{tab:scenario_settings}) to explore the impact of different assumptions and uncertain parameters on the research findings.

In our model, China's domestic aluminum smelters are set to meet the annual primary aluminum demand, defined as the total demand minus the availability of recycled aluminum. This primary demand is a critical factor determining the system-level value of flexibility and the optimal level of strategically retained overcapacity: intuitively, a higher primary demand implies a larger industrial base with greater cumulative potential for seasonal load shifting and peak shaving. We focus on primary aluminum smelters for seasonal flexibility, modeling recycled (secondary) aluminum production as a continuous baseload process. This constraint is driven by three techno-economic factors: (1) unlike the legacy overcapacity in primary smelting, the rapidly expanding recycling sector requires substantial new capital investment, necessitating high utilization rates to amortize marginal capital expenditures; (2) the low energy intensity of recycling (roughly 5\% of primary production) yields electricity cost savings during peak months that are insufficient to offset the depreciation of idle capacity; and (3) the temporal rigidity of scrap supply and its high inventory holding costs render seasonal shutdowns economically unviable. By exploring various demand scenarios, we evaluate how the value of overcapacity evolves as the industry transitions from a primary-dominated era toward a recycling-dominated one.

The flexibility value of aluminum smelting and its resulting operational patterns are governed by the physical constraints of the smelting process, specifically the extent to which individual potlines can adjust their power consumption and the technical difficulty associated with complete shutdowns and restarts. Using performance data from numerical simulations of aluminum smelter operations, as well as existing literature on the shutdown, restart processes, and costs of real-world aluminum smelters, we developed a reduced-order model that represents the key characteristics of aluminum electrolytic cell temperature constraints and restart costs (see Supplementary Note 3 for more details). However, the cost uncertainty is large because there are few historical cases of aluminum smelters being restarted. Thus, we evaluated four comparative scenarios: 

Low: Based on the traditional understanding of the aluminum smelting process, potlines must operate stably with almost no power adjustment capability except for a complete shutdown. Restart costs are based on historical high-cost reports.

Mid (Core Scenario): Based on recent observations in cases such as Yunnan, China, smelters can reduce the operating power of individual potlines by 10\% without significant modifications. Restart costs are also derived from reported values in these Chinese cases.

High: New technologies and methods significantly improve smelter operational flexibility, allowing for power reductions of up to 30\% and substantially lower restart costs. This represents the goal of ongoing research, development, and demonstration efforts to enhance smelter-grid interaction~\citep{lukin_reduction_2016}.

Un-constrained: This is an idealized scenario that disregards the temperature constraints, startup/shutdown constraints, and restart costs of potlines. It only considers rated power constraints, and the value of grid interaction in this scenario represents the theoretical maximum.

The system-level flexibility value of aluminum smelting is also influenced by the costs of energy supply and the availability of alternative flexibility sources. Lower costs for renewable energy or competing flexibility technologies, such as batteries, could diminish the relative value of industrial load flexibility, thereby impacting the economic rationale for retaining overcapacity. To account for this, we consider three main power system technology cost scenarios (see Table \ref{tab:scenario_settings}) that reflect different levels of power generation technology development. All scenarios assume that a variety of competitive, clean, and stable technologies are available, covering a range from high fixed costs and low variable costs to low fixed costs and high variable costs. We also assume the availability of cost-competitive long-duration storage technologies (in the form of hydrogen and hot water) and a certain degree of flexibility in residential heating.

Finally, we introduce a scenario dimension concerning labor and employment flexibility, which may affect the economic feasibility of seasonal operation. We evaluate two comparative cases:
(1) Inflexible labor (core scenario): This represents the most unfavorable case where smelters must maintain their full workforce and associated labor and standby costs regardless of output levels or seasonal shutdowns. Workers are assumed to be unable to transfer to other industries. 
(2) Flexible labor: If a potline is shut down, labor and standby costs are reduced. By modeling combinations of different scenarios for primary aluminum demand, smelter flexibility, technology costs, and labor flexibility, we are able to evaluate the sensitivity of the main power system and aluminum industry outcomes to changes along these dimensions.

\subsection*{Aluminum demand projections}\label{method_projection}

Our objective is to project China's primary aluminum demand for the coming decades, which requires projecting both total aluminum demand and the future supply of recycled aluminum. Our methodology is based on the assumption that per capita aluminum in-use stock tends to saturate as the economy develops, a common approach for projecting the demand for metal raw materials like steel and aluminum~\citep{song_mapping_2020, li_analysis_2022}. To obtain historical aluminum in-use stock data, we sourced historical aluminum production data from the China Statistical Yearbook. We then calculated the aluminum stock by accumulating historical production and accounting for product retirement. Specifically, using China's primary and secondary aluminum production data from 1993 to 2023 and assuming that the lifespan of aluminum products follows a Weibull distribution, we calculated the annual aluminum in-use stock up to 2024.
Next, we input this data into the EnergyPathways (EP) model~\citep{Farid2024}, a mature, open-source software for forecasting commodity demands in energy systems, which has been used in high-profile studies such as the Net-Zero America~\citep{larson2020net} and Net-Zero Australia~\citep{mccall2025topical} projects. The choice of EP is motivated by its proven capability to model future product demand. Using the stock-based saturation method and projected future population from~\citet{chen_provincial_2020}, we used logistic regression (S-curve fitting) to extrapolate future aluminum demand. This approach is well-suited for our study as it captures the decline in primary metal consumption due to slowing population growth, despite continuing economic growth. The EP model projects the annual aluminum stock, total aluminum demand, and retired aluminum products (the available scrap in that year) up to 2060. The calculation formula is as follows:

\begin{equation}
\label{eq_demand}
\text{Stock}_{\text{year}} = \text{Stock}_{\text{year}-1} + \text{Production}_{\text{year}} - \text{Scrap}_{\text{year}}
\end{equation}
Here, $\text{Stock}_{\text{year}}$ represents the extrapolated aluminum stock, $\text{Production}_{\text{year}}$ is the total demand/production for aluminum, and scrap is calculated from the historical stock based on the assumption that product lifespan follows a Weibull distribution. Therefore, the total aluminum demand and available scrap can be calculated from the above formula. Based on the scrap, we can calculate the recycled aluminum production ($\text{Production}_{\text{recycle}} = \text{Scrap} \times \text{Recycle}_{\text{rate}}$). By subtracting the recycled aluminum production from the total aluminum demand for that year, we obtain the primary aluminum demand/production ($\text{Production}_{\text{primary}} = \text{Production} - \text{Production}_{\text{recycle}}$). China is also a major aluminum exporter. Based on current export levels, we have also established two scenarios: reduced exports and increased exports (see Table~\ref{tab:scenario_settings}), with further details provided in Supplementary Note 2.4. Having obtained the annual aluminum demand projections, we assume that the hourly aluminum demand remains constant throughout the year, which simplifies demand distribution yet imposes a stricter supply-demand balance. This demand must be satisfied either through production from aluminum smelters or by drawing from existing aluminum inventories.

\subsection*{Modeling smelter operational flexibility}

This study employs a mixed-integer linear model, an initial version of which was developed by Shen et al.~\citep{shen_improved_2025} as a linear model for production and energy consumption decisions in smelter operations. The model is capable of representing the relationships between internal temperature changes and electrolysis rates in response to applied voltage and current in a potline, as observed in both numerical simulations and experiments~\citep{oye_power_2011,oye_power_nodate}. It is transformed into a linear programming model through piecewise linearization and numerical approximation. To capture the unique flexibility characteristics of aluminum smelters, we introduce binary variables for startup and shutdown decisions, which extends this framework to a mixed-integer linear programming (MILP) model. This extension is essential because aluminum smelters, similar to thermal power units, exhibit discrete operational states and path-dependent costs. Specifically, if a potline remains operational, it must maintain a minimum power level (typically 70--90\% of rated power) to preserve the internal thermal balance of the electrolytic cells. Reducing power further requires a complete shutdown, which then incurs significant restart costs to restore the thermal and chemical equilibrium—costs that can be orders of magnitude higher than those for conventional power units~\citep{smith_analysis_2019}. A pure linear model would fail to capture these all-or-nothing and minimum-stable-output constraints, potentially overestimating the smelter's ability to provide continuous, low-cost load shedding. This enables our model to optimize the smelter's operational decisions over a longer timescale (see Supplementary Note 3 and 5). However, the introduction of integer variables makes it computationally infeasible to directly embed a plant-level operational model into a power system planning model. Therefore, we design a decomposition-based iterative algorithm for solving this problem. To determine the production capacity of smelters, based on the Chinese government's policy that primary aluminum capacity will not exceed 45 million tonnes per year, we assume that the production capacity of aluminum smelters remains unchanged at the 2024 level of 45 million tonnes per year, unless early retirement is chosen. This assumption is based on the typical lifespan of an aluminum smelter, which exceeds 50 years, and the fact that more than 90\% of China's aluminum smelters were built after 2000 (see Supplementary Note 3.3). 

Since not all provinces have complete statistical data on aluminum smelting capacity, we distributed the national capacity to each province based on its 2023 production. We assumed that the capacity share of each province remains unchanged when reducing overcapacity. This approach is justified by the fact that China's aluminum smelting industry currently operates at a high capacity utilization rate of approximately 97\%, meaning that most provinces are running at near full capacity. For estimating the needed workforce for smelters and power plants, we use the average number of workers required per unit of operating capacity, assuming a certain proportion of variable workforce (see Supplementary Note 3.7).

\subsection*{Smelter costing}
The investment, operation, and maintenance costs of aluminum smelters are estimated from annual reports from Chinese smelters~\citep{CHALCO2020AnnualReport}. However, it is challenging to determine the idling and startup/shutdown costs, as there are not many historical cases where smelters were idled for long periods or shut down due to insufficient demand. According to publicly available reports, the normalized cost for restarting potlines can vary from 110,000 CNY/MW to 760,000 CNY/MW~\citep{Hu2009, noauthor_high_2020,noauthor_alcoa_2021}, which is 1000 times higher than restarting coal-fired units~\citep{smith_analysis_2019}. This study does not attempt to predict the future trajectory of smelter restart costs, but rather presents extreme cases in addition to the baseline case to represent the two extremes of the spectrum for the development of flexible aluminum smelting technologies, resulting in a total of four smelter flexibility cases. In the Low-flexibility and Mid-flexibility cases, the smelter flexibility parameters are set based on current literature and reports. The former's restart cost is estimated at 760,000 CNY/MW in a case of a U.S.-owned smelter. The latter's restart cost is from a documented restart event at a smelter in Yunnan, China, estimated at 110,000 CNY/MW. 
For smelters to conduct seasonal production and help adapt to fluctuations in the power grid's supply and demand, they may need to increase the storage period of aluminum products to act as a buffer for balancing production and demand. Therefore, we also need to model the warehouse costs of aluminum products. Product warehouse costs are estimated by multiplying the actual short-term warehouse rental fees in each region for 2023 by the required warehouse area~\citep{Colliers2024Logistics}. However, this may lead to an overestimation, as long-term rental fees are typically lower. A more detailed description of the smelter operational cost calculation method is in Supplementary Note 3.3.

\subsection*{Power system capacity expansion model}
This study evaluates the value of industrial load flexibility within a national-scale, multi-period capacity expansion framework. A provincial resolution is needed because China's aluminum industry is unevenly distributed across multiple provinces, and the highly interconnected nature of China's power grid means that the impact of this flexibility is likely to span across provincial boundaries. The modeling aims to identify the least-cost infrastructure investment and operational dispatch decisions needed to meet climate targets. While China's national target is to achieve carbon neutrality by 2060~\citep{yang_chinas_2024}, we define a sector-specific 2050 net-zero scenario for the power and heating sectors (DP-2050). This is because easier-to-abate sectors may be decarbonized earlier in order to achieve economy-wide net-zero emissions by 2060. Power and residential heating are relatively easy-to-abate sectors due to the rapid decline in renewable energy costs and the maturity of technologies such as heat pumps, and thus achieving net-zero in power and heating sectors by 2050 may be a necessary prerequisite for the decarbonization of other end-use sectors. Furthermore, according to several studies and scenario analyses, China's power sector is expected to reach net-zero emissions between 2045 and 2055~\citep{liu_chinas_2025}, allowing the sector to transition toward net-negative emissions (e.g., via bioenergy with carbon capture and storage) to offset residual emissions from hard-to-abate sectors by 2060. This timeline is supported by studies from the International Energy Agency~\citep{iea_roadmap_2021}, Tsinghua University~\citep{tsinghua_iccsd_2020}, and the Rocky Mountain Institute~\citep{rmi_china_2050}. To compare the effects of different carbon reduction constraints, we set up scenarios with 20\%, 60\%, and 100\% reductions in carbon emissions for 2030, 2040, and 2050, respectively, based on 2020 emission levels. 

We utilize the Python for Power System Analysis (PyPSA) framework, a widely-recognized, open-source tool for high-resolution energy system optimization. PyPSA has been extensively applied in numerous peer-reviewed studies to explore large-scale decarbonization pathways, most notably within the European context through the PyPSA-Eur project~\citep{PyPSAEur}. Unlike copper-plate models, PyPSA captures the essential physics of transmission through a linearized DC approximation of power flow, enforcing Kirchhoff's laws and identifying network bottlenecks that drive storage requirements and renewable curtailment. Furthermore, PyPSA natively handles sector coupling, allowing electricity, heat, hydrogen, and gas to be co-optimized. In this study, we utilize PyPSA-China, a peer-reviewed model build on PyPSA that incorporates the specialized dataset for China's net-zero transition analysis~\citep{zhou_multienergy_2024}.

\subsection*{Dimensionality reduction and tailored iterative solution}
Our objective is to address the computational intractability that arises when integrating detailed smelter operation models (which include integer variables) into a large-scale power system planning model. We adopt an iterative decomposition framework based on the Data-Driven Dimension Reduction (D3R) method proposed by Lyu et al.~\citep{lyu_data-driven_2025}. D3R utilizes inverse optimization to learn the feasible operating region of complex industrial loads. It trains a low-dimensional polytope that best approximates the smelter's physical flexibility (including thermal and stoichiometry constraints) while remaining compatible with linear optimization. The D3R framework has been rigorously verified for its ability to maintain high approximation accuracy while reducing the computational burden by orders of magnitude, making it uniquely suitable for modeling the flexibility potential of energy-intensive industries in large-scale energy systems~\citep{lyu_data-driven_2025}.

The solution process begins by embedding this linearized smelter model obtained through the D3R framework into the electricity system capacity expansion model. Solving this initial integrated model yields a first estimate of electricity system capacity expansion results and the hourly local marginal prices (LMPs) for electricity in each province.
Then, these LMPs are fixed and used as an exogenous price signal for the detailed potline-level smelter operation models. This step allows each smelter's 8760-hourly cost-minimization problem (including all unit commitment constraints) to be solved independently and in parallel. Once the detailed optimal production schedule and therefore the 8760-hourly load profile is determined for each smelter, we fix these schedules in the main capacity expansion model and then re-optimize all remaining power system expansion and dispatch variables. This re-optimization generates a new, updated set of LMPs, which are fed back into the detailed smelter models. This entire process is iterated until the objective function of the planning model converges. For more details, see Supplementary Note 5.

\subsection*{Limitations and opportunities}

Our study has several limitations that offer avenues for future research. Regarding smelter representation, our model aggregates plants at the provincial level (see Supplementary Note 5.2 for details) and, due to computational constraints, assumes fixed capacity ratios between provinces. This spatial aggregation overlooks technological heterogeneity among plants. The same capacity retaining ratio across the nation prevents us from deriving an optimized, facility-level retirement strategy. In reality, a more granular representation of retirement decisions would likely favor retaining plants that are more modern and technically efficient, which would be better equipped to provide grid flexibility. Future work could leverage efficient frameworks to optimize the spatial distribution of smelter retirement or migration, thereby capturing regional flexibility value, or, with enough data, offer plant-level operational strategies. We also do not differentiate between smelting technologies, retrofitting options, or stoichiometry of aluminum smelting~\citep{shen_technologies_2024, tan_different_2025}. Aluminum smelting is the most electrified energy-intensive industry and is uniquely suited for seasonal stockpiling due to its homogeneity and negligible storage degradation, whereas other commodities (e.g., steel, cement) face higher physical risks and holding costs. When extending this framework to other heavy industries, future research must incorporate sector-specific storage degradation and holding costs. Subsequent research could explore the co-optimization of multiple energy-intensive sectors to develop a more general understanding of how other energy-intensive industries can collectively manage overcapacity and navigate the low-carbon transition. More broadly, the question of how to leverage industrial-energy sector coupling to mitigate the socio-economic impacts of energy decarbonization and industrial restructuring represents a promising future research direction.

\section*{Data availability}

All PyPSA input and results datasets relevant to this study are available via Zenodo at \url{https://doi.org/10.5281/zenodo.19600477}\citep{lyu2026industrial_overcapacity_supplementary}. Additional data are available from the corresponding author on reasonable request.

\section*{Code availability}

The PyPSA Power System Analysis model is open source and available via GitHub at https://github.com/PyPSA/PyPSA. Source code for the modified version of PyPSA-China used in this work is available via Github at https://github.com/Rick10119/PyPSA-China-THU.

\section*{Acknowledgment}
This work was supported by National Natural Science Foundation of China (No. 52130702, No. 72342007, No. 72422015) and Carbon Neutrality and Energy System Transformation project initiated by Tsinghua University.
The simulations presented in this article were performed on the Della cluster at Princeton University, which is managed by the Princeton Institute for Computational Science and Engineering (PICSciE) and the Office of Information Technology's Research Computing department.
The authors thank Tengmu Li for perspectives on the seasonal rhythms of the contemporary workforce, Zongzhi Du for providing consultation on aluminum corporations in China, and Eric Larson for providing feedback on the manuscript.

\section*{Author contributions}
R.L., H.G., C.K. and J.D.J. conceptualized the study. R.L. and J.D.J. developed the experimental design. R.L., Y.S. and H.G. designed and performed the smelter simulations. R.L. developed the optimization model and other input datasets. R.L. performed the formal analysis, visualization and investigation, and produced the figures. R.L. drafted, revised and finalized the manuscript. J.D.J., A.L., H.L., H.G., J.W., E.D. and C.K. advised on the analysis and reviewed and revised the manuscript.

\section*{Competing interests}

J.D.J. is a co-founder and Chief Technology Officer of Firma Power, LLC, which develops optimization and decision support software for energy systems. J.D.J. is part owner of DeSolve, LLC, which provides techno-economic analysis and decision support for clean energy technology ventures and investors. A list of clients can be found at https://www.linkedin.com/in/jessedjenkins. He serves on the advisory boards of Eavor Technologies Inc., a closed-loop geothermal technology company; Rondo Energy, a provider of high-temperature thermal energy storage and industrial decarbonization solutions; Dig Energy, a developer of low-cost drilling solutions for ground-source heat pumps; Karman Industries; and Emerald AI and has an equity interest in each company. He also serves as a technical advisor to MUUS Climate Partners and Energy Impact Partners, both investors in early-stage climate technology companies.


\bibliographystyle{elsarticle-num-names}
\bibliography{reference}

\end{document}